\documentclass[twocolumn,showpacs,preprintnumbers,amsmath,amssymb]{revtex4}
 
\usepackage{graphicx}
\usepackage{dcolumn}
\usepackage{bm}
 
\newcommand{\pro}[2]{\langle{#1}|{#2}\rangle}

\newcommand{\bra}[1]{\langle{#1}|}
\newcommand{\ket}[1]{|{#1}\rangle}

\newcommand{\dgg}{^{\dagger}}
\newcommand{\Tr}{{\rm Tr}\hspace{0.07cm}}

\newcommand{\im}{{\rm i}}

\newcommand{\half}{\frac{1}{2}}

\begin{document}
 
\preprint{APS/123-QED}
 
\title{Parametrization of the feedback Hamiltonian realizing a 
pure steady state}
 
\author{Naoki Yamamoto}
 \email{naoki@cds.caltech.edu}
\affiliation{%
Control and Dynamical Systems, California 
Institute of Technology, Pasadena, California 91125, USA
}%
\date{\today}

\begin{abstract}
Feedback control is expected to considerably protect quantum states against 
decoherence caused by interaction between the system and environment. 
Especially, Markovian feedback scheme developed by Wiseman 
can modify the properties of decoherence and eventually recover the 
purity of the steadystate of the corresponding master equation. 
This paper provides a condition for which the modified master equation 
has a pure steady state. 
By applying this condition to a two-qubit system, we obtain a complete 
parametrization of the feedback Hamiltonian such that the steady state 
becomes a maximally entangled state. 
\end{abstract}
 
\maketitle


\section{Introduction}

The time evolution of a quantum state under Markovian decoherence 
is usually modeled by a master equation in the Lindblad form 
\cite{lindblad}, 
\begin{eqnarray}
& & \hspace*{-1em}
\label{master-eq}
    \frac{d\rho}{dt}={\cal L}\rho
                    =-\im[H, \rho]+\sum_{k=1}^{M}{\cal D}[b_k]\rho, 
\\ & & \hspace*{1em}
    {\cal D}[b_k]\rho:=b_k\rho\hspace*{0.1em}b_k\dgg
                         -\half b_k\dgg b_k\hspace*{0.1em}\rho
                         -\half\rho\hspace*{0.1em}b_k\dgg b_k, 
\nonumber
\end{eqnarray}
where $H=H\dgg$ denotes a driving Hamiltonian, and $b_k$ are the Lindblad 
operators that generate decoherence. 
The dynamics (\ref{master-eq}) often reduces the purity of $\rho$, where 
the purity is defined by $p:=\Tr(\rho^2)$. 
For example, a master equation that preserves the identity operator 
is strictly purity decreasing. 
(Conversely, any finite-dimensional purity-decreasing dynamics preserves 
the identity operator \cite{lidar0}.) 
Since loss of purity is the most critical issue in many quantum information 
technologies \cite{nielsen}, we must exploit effective methodologies to 
prepare desirable pure states even under decoherence. 

Quantum control shows a high possibility to overcome the problems of 
decoherence. 
In particular, quantum filtering theory, which was pioneered by Belavkin 
\cite{belavkin1,belavkin2}, enables us to feed measurement data continuously 
back in order to control quantum systems. 
Actually, in the framework of Markovian feedback proposed by Wiseman 
\cite{wiseman1}, a data-dependent feedback Hamiltonian is continuously 
added to the system, and the structure of the corresponding master 
equation is modified. 
Markovian feedback method has been widely used for several types of quantum 
systems. 
For example, a master equation describing a two-level atom was modified by 
the feedback so that any pure state becomes a steady state of the 
dynamics \cite{wang1}. 
Further, in the case of a system consisting of two two-level atoms (qubits) 
coupled to a single-mode cavity field that is heavily damped, it was proved 
that the entanglement of the steady state was increased by the application 
of Markovian feedback \cite{wang2}. 

However, the selection of the data-dependent feedback Hamiltonian is 
usually based on intuitive observations. 
In other words, there are no systematic guidelines to modify the master 
equation to a desirable one. 
Actually, the above-mentioned steady state of the controlled two-qubit 
system was not a maximally entangled state but a mixed state. 

Therefore, for finite-dimensional quantum systems, this paper 
characterizes a feedback Hamiltonian completely such that the modified 
master equation has a pure steady state. 
By applying this condition to the two-qubit system, we obtain a 
parametrization of the feedback Hamiltonian that causes the steady state 
to be not only a pure state but also a maximally entangled state.


\section{A characterization of the feedback Hamiltonian}

\subsection{The condition for a pure steady state}

The steady state of the master equation (\ref{master-eq}) is 
a quantum state $\rho_o$ satisfying ${\cal L}\rho_o=0$. 
Remarkably, the steady state $\rho_o$ is stable in the following sense. 
Let us define the matrix elements $\rho_{ij}~(1\leq i,j\leq n)$ 
of an $n$-dimensional state $\rho$ as follows. 
Set $\rho_{ii}:=\nu_i,~\nu_i\in{\mathbb R}$ for $1\leq i\leq n-1$, which leads 
to $\rho_{nn}=1-\sum_{i}\nu_i$. 
For $i>j$ set $\rho_{ij}:=\lambda_{ij}+\im\mu_{ij}$ with 
$\lambda_{ij},~\mu_{ij}\in{\mathbb R}$. 
For $i<j$ set $\rho_{ij}=\rho_{ji}^*$. 
Collecting the real numbers $\nu_i,~\lambda_{ij}$, and $\mu_{ij}$ into 
a vector $x\in{\mathbb R}^{n^2-1}$, we can describe the master equation as 
an affine equation $\dot{x}=Ax+a$, where the matrix $A$ and the vector $a$ 
are uniquely determined by the above rule specifying $x$. 
Then, all the eigenvalues of $A$ have non-positive real parts. 
This implies that for an initial state $\rho_o+\delta\rho$ 
($\delta\rho$ implies a small perturbation from $\rho_o$), we can always 
find $\rho(t)$ sufficiently close to $\rho_o$ for all $t>0$. 
Especially in the case when all the eigenvalues of $A$ have negative real 
parts, $\rho_o$ becomes a unique steady state corresponding to 
$x_o=-A^{-1}a$ in the vector form; eventually every trajectory that obeys 
the dynamics must converge into $\rho_o$. 

By combining the above discussion with the fact that pure states are the 
basis of many quantum information technologies, we aim to characterize 
a class of the Lindblad operators $b_k$ and the Hamiltonian $H$ such that 
the master equation (\ref{master-eq}) has a pure steady state. 

{\bf Theorem 1.}~
{\it 
The master equation (\ref{master-eq}) has a pure steady state if and 
only if $b_{k}~(k=1,\ldots,M)$ and $\im H+(1/2)\sum_{k}b_{k}\dgg b_{k}$ 
have a common eigenvector $\ket{\phi}$. 
Then, the steady state is given by $\rho=\ket{\phi}\bra{\phi}$. 
}

{\bf Proof:}~
Substituting $\rho=\ket{\phi}\bra{\phi}$ into the equation ${\cal L}\rho=0$ 
and multiplying by $\ket{\phi}$ on both sides, we have 
\begin{equation}
\label{pure-equal}
    \sum_{k=1}^{M}\Big[
      |\bra{\phi}b_{k}\ket{\phi}|^{2}-\bra{\phi}b_{k}\dgg b_{k}\ket{\phi}
            \Big]=0. 
\end{equation}
Since the Schwarz inequality yields 
\[
    \bra{\phi}b_{k}\dgg b_{k}\ket{\phi}\pro{\phi}{\phi}
       \geq |\bra{\phi}b_{k}\ket{\phi}|^{2}, 
\]
Eq. (\ref{pure-equal}) is attained if and only if 
$b_{k}\ket{\phi}=\mu_{k}\ket{\phi}~(k=1,\ldots,M)$ holds for some 
$\mu_{k}\in{\mathbb C}$. 
Thus, owing to $b_k\ket{\phi}\bra{\phi}b_k\dgg=|\mu_k|^2\ket{\phi}\bra{\phi}$, 
the equation ${\cal L}\ket{\phi}\bra{\phi}=0$ is transformed to 
\begin{equation}
\label{F-eq}
    F\ket{\phi}\bra{\phi}+\ket{\phi}\bra{\phi}F\dgg=0, 
\end{equation}
where 
\[
     F:=-\im H
        +\sum_{k=1}^{M}\Big(\half|\mu_{k}|^{2}-\half b_{k}\dgg b_{k}\Big). 
\]
By denoting the $k$th elements of $\ket{\phi}$ and $F\ket{\phi}$ by 
$\phi_{k}$ and $f_{k}$, respectively, we observe that the $(j, k)$ element 
of the matrix equation (\ref{F-eq}) is 
$f_{j}\phi_{k}^{*}+\phi_{j}f_{k}^{*}=0$, which becomes 
\begin{equation}
\label{fracEq}
    \Big(\frac{f_{j}}{\phi_{j}}\Big)+\Big(\frac{f_{k}}{\phi_{k}}\Big)^{*}=0, 
\end{equation}
when $\phi_{j}\neq 0$ and $\phi_{k}\neq 0$. 
Then, $\beta_{j}:=f_{j}/\phi_{j}$ satisfies $\beta_{j}+\beta_{j}^{*}=0$. 
Thus, $\beta_{j}$ is a pure imaginary number: 
$\beta_{j}=\im\beta'_{j}, \hspace*{0,3em} \beta'_{j}\in{\mathbb R}$. 
Therefore, Eq. (\ref{fracEq}) yields $\beta'_{j}-\beta'_{k}=0$. 
This leads to the conclusion that $\beta_{j}$ is independent of the 
index $j$. 
Accordingly, we obtain $f_{j}=\beta\phi_{j}$ by defining a constant 
$\beta:=\beta_{j},~^{\forall}j$. 
When $\phi_{j}=0$, it immediately yields $f_{j}=0$. 
Consequently, we obtain $F\ket{\phi}=\beta\ket{\phi}$. 
It is easy to verify the ``if" condition. 
\\
\\
The proof of Theorem 1 is very similar to the one found in \cite{lidar1}: 
the theory of {\it decoherence-free subspace} (DFS) 
\cite{lidar1,lidar2,lidar3}. 
We now consider a relationship between DFS and a pure steady state. 
Let the system Hilbert space ${\cal H}$ decompose into a direct sum as 
${\cal H}={\cal H}_{{\rm df}}\oplus{\cal H}_{{\rm df}}^{\perp}$, and 
partition the system state, the Hamiltonian, and the Lindblad operators 
into blocks as follows: 
\[
    \rho=\left[ \begin{array}{cc}
            \rho_1 & \rho_2 \\
            \rho_2\dgg & \rho_3 \\
         \end{array} \right],~~
    H=\left[ \begin{array}{cc}
            H_1 & H_2 \\
            H_2\dgg & H_3 \\
         \end{array} \right],~~
    b_k=\left[ \begin{array}{cc}
            P_k & Q_k \\
            R_k & S_k \\
         \end{array} \right]. 
\]
Then, ${\cal H}_{{\rm df}}$ is called decoherence free if and only if 
$\rho_1$ undergoes $\dot{\rho}_1=-\im [H_1,\rho_1]$. 
When we assume $\rho_2(0)=\rho_3(0)=0$, the necessary and sufficient 
condition for ${\cal H}_{{\rm df}}$ to be decoherence free is that 
the Lindblad operators and the Hamiltonian satisfy 
\begin{equation}
\label{dfs-condition}
    P_k=\alpha_k I,~~
    R_k=O,~~
    H_2+\frac{\im}{2}\sum_{k}\alpha_k^* Q_k=O, 
\end{equation}
where $\alpha_k$ are arbitrary complex scalars \cite{lidar3}. 
(If $\rho_2(0)$ and/or $\rho_3(0)$ do not vanish, in addition to 
(\ref{dfs-condition}), $Q_k=O$ has to hold.) 
We then observe that the DFS always includes pure steady states. 
Actually, a pure state $\ket{\psi}=\ket{h}\oplus\ket{0}
\in{\cal H}_{{\rm df}}\oplus{\cal H}_{{\rm df}}^{\perp}$, where 
$\ket{h}$ is an eigenvector of $H_1$ and $\ket{0}$ is the zero vector, 
is a steady state of the dynamics because the unitary evolution 
$\dot{\rho}_1=-\im [H_1,\rho_1]$ has a steady state 
$\rho_1=\ket{h}\bra{h}$. 
It is also observed that the matrix $\im H+(1/2)\sum_k b_k\dgg b_k$ is now 
represented by 
\[
      \left[ \begin{array}{cc}
       \im H_1 +\half\sum_k|\alpha_k|^2 
                    & \im H_2 +\half\sum_k\alpha_k^* Q_k \\
       O & \im H_3 +\half\sum_k(Q_k\dgg Q_k+R_k\dgg R_k) \\
         \end{array} \right], 
\]
and thus, the condition of Theorem 1 indeed holds; we have 
$b_k\ket{\psi}=\alpha_k\ket{\psi}$ and 
\[
    \Big[\im H+\half\sum_k b_k\dgg b_k\Big]\ket{\psi}
        =\Big[\im h+\half\sum_k |\alpha_k|^2 \Big]\ket{\psi}, 
\]
where $h\in{\mathbb R}$ denotes the eigenvalue of $H_1$ corresponding to 
$\ket{h}$. 
The above discussion indicates that the condition for a DFS is more 
general than that for a pure steady state.


\subsection{The modified master equation via feedback}

The dynamical evolution of a quantum state under homodyne measurement 
is described by the {\it stochastic master equation} \cite{belavkin1,
belavkin2,wiseman1}, 
\begin{eqnarray}
& & \hspace*{-1em}
    d\rho_c=-\im[H, \rho_c]dt+{\cal D}[c]\rho_c\hspace*{0.1em}dt
                             +{\cal H}[c]\rho_c\hspace*{0.1em}dW, 
\nonumber \\ & & \hspace*{-1em}
    dy=\Tr[(c+c\dgg)\rho_c]dt+dW, 
\nonumber
\end{eqnarray}
where $\rho_c$ denotes the quantum state conditioned on the 
measurement data $y(t)$, and $dW$ is the standard Wiener increment with 
mean zero and variance $dt$. 
The superoperator ${\cal H}[c]\rho:=c\rho+\rho c\dgg-\Tr[(c+c\dgg)\rho]\rho$ 
represents the stochastic jump associated with the continuous measurement. 
It must be noted that we consider only one decoherence term 
${\cal D}[c]\rho_c$ here, which is caused by the interaction between the 
system and measurement apparatus. 

In the theory of Markovian feedback, adding a data-dependent feedback 
Hamiltonian $I(t)F=(dy/dt)F$, we can modify the properties of the decoherence 
as follows: 
The infinitesimal increment of the conditioned quantum state is 
given by $\rho_c(t+dt)={\rm e}^{I(t)dt{\cal K}}[\rho_c(t)+d\rho_c(t)]$, 
where ${\cal K}\rho=-\im[F,\rho]$. 
This leads to the following modified stochastic master equation: 
\begin{eqnarray}
& & \hspace*{-1em}
    d\rho_c=-\im[H, \rho_c]dt+{\cal D}[c]\rho_c\hspace*{0.1em}dt
                             +{\cal H}[c]\rho_c\hspace*{0.1em}dW
\nonumber \\ & & \hspace*{1.8em}
    -\im[F, c\rho_c+\rho_c c\dgg]dt
           +{\cal D}[F]\rho_c\hspace*{0.1em}dt-\im[F, \rho_c]dW. 
\nonumber
\end{eqnarray}
The controlled master equation is obtained by simply dropping the 
stochastic term in the above equation as follows: 
\begin{equation}
\label{feedback-eq}
    \frac{d\rho}{dt}
         =-\im\Big[H+\half c\dgg F+\half Fc,\hspace{0.09cm}\rho\Big]
                     +{\cal D}[c-\im F]\rho. 
\end{equation}
Hence from Theorem 1, a pure state $\ket{\phi}$ becomes a steady state 
of the modified master equation (\ref{feedback-eq}) if and only if 
the two matrices $\im H'+(1/2)c'\mbox{}\dgg c'$ and $c'$ have a common 
eigenvector $\ket{\phi}$, where $H':=H+(1/2)c\dgg F+(1/2)Fc$ and 
$c':=c-\im F$. 
A direct calculation yields the following objective condition. 

{\bf Theorem 2.}~
{\it 
The modified master equation (\ref{feedback-eq}) has a pure steady state 
if and only if two matrices 
\[
     A=\im H+\im Fc+\half c\dgg c+\half F^2,~~B=c-\im F
\]
have a common eigenvector $\ket{\phi}$. 
Then, the steady state is given by $\rho=\ket{\phi}\bra{\phi}$. 
}


\section{Examples}

\subsection{Purification of a single atom}

As a simple example, let us reconsider the control problem for a single 
two-level atom \cite{wang1}. 
The Lindblad operator given by $\frac{\sqrt{\gamma}}{2}\sigma_{-}
=\frac{\sqrt{\gamma}}{2}(\sigma_x-\im\sigma_y)$, where 
$\sigma_i~(i=x,y,z)$ denote Pauli matrices, 
represents spontaneous emission. 
The feedback Hamiltonian and driving Hamiltonian are given by 
$F=\lambda\sigma_y$ and $H=\alpha\sigma_y$, respectively, where 
$\lambda\in{\mathbb R}$ and $\alpha\in{\mathbb R}$ are controllable 
parameters. 
Without the feedback control (i.e., $\lambda=0$), only the ground state 
$\ket{g}=[0~1]^T$ can be pure when $\alpha=0$. 
Let us compute the condition of the control parameters such that 
the modified master equation has a pure steady state. 
Since $\dim{\cal H}=2$, the condition in Theorem 2 is equivalent to 
${\rm det}[A,B]=0$, which leads to 
\[
     \alpha^2+\Big[
           \Big(\lambda+\frac{\sqrt{\gamma}}{2}\Big)^2-\frac{\gamma}{8}
              \Big]^2=\Big(\frac{\gamma}{8}\Big)^2. 
\]
Hence, $\alpha$ and $\lambda$ must satisfy the relations 
\begin{equation}
\label{single-atom}
    \alpha=\frac{\gamma}{4}\sin\theta\cos\theta,~~
    \lambda=-\frac{\sqrt{\gamma}}{2}(1\pm\cos\theta), 
\end{equation}
where $\theta\in{\mathbb R}$ is a real parameter. 
Then we have steady states 
\[
    \ket{\phi_+}=\left[ \begin{array}{c}
                        \cos(\theta/2) \\
                        \pm\sin(\theta/2) \\
                     \end{array} \right],~~
    \ket{\phi_-}=\left[ \begin{array}{c}
                        \sin(\theta/2) \\
                        \pm\cos(\theta/2) \\
                     \end{array} \right], 
\]
where $\phi_{\pm}$ corresponds to the sign in $\lambda$. 
It should be noted that Eq. (\ref{single-atom}) is a necessary and 
sufficient condition for the master equation to have a pure steady state, 
whereas it was derived as only a necessary condition in \cite{wang1}.


\subsection{Dynamical creation of a maximally entangled state}

The study \cite{wang2} successfully applied the Markovian feedback method 
to a two-qubit system for the following setup. 
The Lindblad operator associated with the measurement is
\[
     c=-\im\sqrt{\gamma}(\sigma_{-}\otimes I+I\otimes\sigma_{-})
      =-2\im\sqrt{\gamma}
         \left[ \begin{array}{cccc}
          0 & 0 & 0 & 0 \\
          1 & 0 & 0 & 0 \\
          1 & 0 & 0 & 0 \\
          0 & 1 & 1 & 0 \\
         \end{array} \right]. 
\]
The feedback and the driving Hamiltonian are given by 
\begin{equation}
\label{original}
   F_o=\lambda J_x,~~
   H_o=\alpha J_x,
\end{equation}
where $\lambda$ and $\alpha\in{\mathbb R}$ are the controllable parameters, 
and $J_x:=(\sigma_x\otimes I+I\otimes\sigma_x)/2$. 
Then, it was shown that the entanglement, which is measured by the 
so-called ``concurrence" \cite{wootters}, of the steady state of the 
modified master equation was larger than that without feedback. 
However, the steady state was a mixed state for any of the parameters 
$\alpha$ and $\lambda$. 

Therefore, assuming that we can make any kind of feedback Hamiltonian $F$, 
our aim is to characterize $F$ such that the maximally entangled state 
$\ket{\Phi}=(1/\sqrt{2})(\ket{00}+\ket{11})$ becomes a steady state of 
the master equation. 
The driving Hamiltonian is given by $H_o$ in Eq. (\ref{original}). 
First, from the condition $B\ket{\Phi}=(c-\im F)\ket{\Phi}=k\ket{\Phi}$ 
in Theorem 2, the feedback Hamiltonian $F=\{f_{ij}\}$ must satisfy 
\begin{eqnarray}
\label{element-condition}
& & \hspace*{-1em}
    f_{11}=f_{44},~~
    f_{34}=-f_{13}^*-2\sqrt{\gamma},
\nonumber \\ & & \hspace*{-1em}
    f_{24}=-f_{12}^*-2\sqrt{\gamma},~~
    f_{14}=-\mu-f_{11}, 
\end{eqnarray}
where we have defined $\mu:=-\im k\in{\mathbb R}$. 
Due to the relations (\ref{element-condition}), the condition 
$A\ket{\Phi}=\nu\ket{\Phi}$ in Theorem 2 becomes 
\begin{equation}
\label{4-entries}
      \left[ \begin{array}{c}
          \sqrt{\gamma}(f_{12}+f_{13})+4\gamma+\mu^2/2   \\
          \im\alpha+\sqrt{\gamma}(f_{22}+f_{23})+\sqrt{\gamma}\mu   \\
          \im\alpha+\sqrt{\gamma}(f_{33}+f_{23}^*)+\sqrt{\gamma}\mu   \\
          -\sqrt{\gamma}(f_{12}+f_{13})-4\gamma+\mu^2/2   \\
         \end{array} \right]
      =\nu\left[ \begin{array}{c}
                        1 \\
                        0 \\
                        0 \\
                        1 \\
                     \end{array} \right]. 
\end{equation}
Thus, from the second and third entries, we obtain 
\begin{eqnarray}
& & \hspace*{-1em}
    2\im\alpha+\sqrt{\gamma}(f_{22}+f_{33}+f_{23}+f_{23}^*)
              +2\sqrt{\gamma}\mu=0,
\nonumber \\ & & \hspace*{-1em}
    (f_{22}-f_{33})+(f_{23}-f_{23}^*)=0,
\nonumber
\end{eqnarray}
which implies that $\alpha=0,~f_{22}=f_{33}$, and $f_{23}\in{\mathbb R}$. 
Accordingly, we have $f_{23}=-f_{22}-\mu$. 
The first and fourth entries in Eq. (\ref{4-entries}) yield 
$f_{12}+f_{13}+4\sqrt{\gamma}=0$ and $\mu^2=2\nu$. 
As a result, the feedback Hamiltonian that enables $\ket{\Phi}$ to become 
a steady state is completely parametrized by 
\begin{eqnarray}
& & \hspace*{-1em}
\label{hamil-F}
    F=(x_1+x_2)I\otimes I-(\mu+x_1+x_2)\sigma_x\otimes\sigma_x
\nonumber \\ & & \hspace*{1em}
    \mbox{}+(x_1-x_2)(\sigma_y\otimes\sigma_y+\sigma_z\otimes\sigma_z)
\nonumber \\ & & \hspace*{1em}
    \mbox{}+(x_3+\sqrt{\gamma})(I\otimes\sigma_x)
              -(x_3+3\sqrt{\gamma})(\sigma_x\otimes I)
\nonumber \\ & & \hspace*{1em}
    \mbox{}+x_4(\sigma_y\otimes\sigma_z-\sigma_z\otimes\sigma_y)
\nonumber \\ & & \hspace*{1em}
    \mbox{}-\sqrt{\gamma}(\sigma_x\otimes\sigma_z+\sigma_z\otimes\sigma_x), 
\end{eqnarray}
where $x_i~(i=1,\ldots,4)$ are real parameters. 
Especially when $x_1=x_2=x_4=\mu=0$ and $x_3=-2\sqrt{\gamma}$, 
the Hamiltonian (\ref{hamil-F}) reduces to 
\begin{eqnarray}
& & \hspace*{-1em}
    F=-\sqrt{\gamma}(\sigma_x\otimes I+I\otimes\sigma_x)
\nonumber \\ & & \hspace*{1.1em}
    -\sqrt{\gamma}(\sigma_x\otimes\sigma_z+\sigma_z\otimes\sigma_x). 
\nonumber
\end{eqnarray}
Therefore, we need to add the global Hamiltonian represented by the 
second term to the local Hamiltonian $F_o$.


\section{Concluding remarks}

Although we could have found feedback Hamiltonians in the two typical 
examples such that the modified master equations have a pure 
steady state, the applicable cases are limited. 
In fact, we cannot find such a convenient Hamiltonian in the following two 
cases even under some ideal assumptions of the feedback scheme (e.g., 
negligible time delay of the feedback). 
The first is the case where there are uncontrollable decoherence 
effects $b_k$ and they have no common eigenvector. 
In this case, obviously, we cannot find a common eigenvector of $b_k$ and 
$c-\im F$ for any Hamiltonian $F$. 
The second case is as follows: 
If the measurement efficiency, denoted by $\eta$, is less than $1$, 
the modified master equation takes the form \cite{thomsen} 
\begin{eqnarray}
\label{inefficient}
& & \hspace*{-1em}
      \frac{d\rho}{dt}=-\im\Big[H+\half c\dgg F+\half Fc,~\rho\Big]
                     +{\cal D}[c-\im F]\rho
\nonumber \\ & & \hspace*{3em}
          \mbox{}+{\cal D}\Big[\sqrt{\frac{1-\eta}{\eta}}F\Big]\rho. 
\end{eqnarray}
From Theorem 1, the dynamics (\ref{inefficient}) has a pure 
steady state $\ket{\phi}$ if and only if $\ket{\phi}$ is an eigenvector 
of $F$; this directly indicates that $\ket{\phi}$ is a common 
eigenvector of $c$ and $\im H+(1/2)c\dgg c$. 
Hence, if $c$ and $\im H+(1/2)c\dgg c$ do not share a common eigenvector, 
we are unable to achieve our objective. 
In other words, the feedback does not have any ability to produce 
a steady state. 
This leads us to conclude that we must exploit basic designing 
methods of a feedback Hamiltonian in order to obtain the purest possible 
mixed steady state. 
%
%
%
%
%

\end{document}